# New Experimental limit on Optical Photon Coupling to Neutral, Scalar Bosons

A. Afanasev,[1]  O.K. Baker,[2]  K.B. Beard,[3]  G. Biallas,[4]  J. Boyce,[4]  M. Minarni,[5]
R. Ramdon,[1]  M. Shinn,[4]  P. Slocum[2]

[1] Department of Physics, Hampton University, Hampton, VA 23668, USA
[2] Department of Physics, Yale University, P.O.Box 208120, New Haven, CT 06520, USA
[3] Muons, Inc., 552 N. Batavia Avenue, Batavia, IL 60510, USA
[4] Free Electron Laser Division, Jefferson Laboratory, 12000 Jefferson Avenue, Newport News, VA 23606, USA
[5] Department of Physics, Riau University (UNRI), Fisika FMIPA UNRI, Kampus Binawidya km 12.5 Simpang Baru, Pekanbaru, Riau 28293, Indonesia

We report on the first results of a sensitive search for scalar coupling of photons to a light neutral boson in the mass range of approximately 1.0 milli-electron volts and coupling strength greater than $10^{-6}$ GeV$^{-1}$ using optical photons. This was a photon regeneration experiment using the "light shining through a wall" technique in which laser light was passed through a strong magnetic field upstream of an optical beam dump; regenerated laser light was then searched for downstream of a second magnetic field region optically shielded from the former. Our results show no evidence for scalar coupling in this region of parameter space.

Several theories in particle physics as well as cosmology predict the existence of at least one scalar, that is, spin-zero, boson [1-9]. Many theories of physics beyond the SM (BSM) can accommodate scalars with much smaller masses and weak couplings to SM fields [10-12]. For the latter, there is renewed interest in experimental searches for sub-electron volt mass spin-zero, weakly interacting particles, triggered in large part by the recent PVLAS collaboration claims [13], now disclaimed [14], of an anomalous rotation of polarized laser light when it propagates through a magnetic field. The experimental programs that explore the parameter space of weakly interacting, light, spin-zero, bosons by and large all use the "light shining through a wall" (LSW) technique of photon regeneration [15]: laser photons are sent through a strong magnetic field where some of them can convert into low-mass, weakly interacting bosons via the Primakoff effect, these bosons then pass through a wall that serves to block the incident laser light, and reconvert into photons in a second magnetic field in a similar manner, as shown in Fig. 1.

The Light Pseudoscalar and Scalar Particle Search (LIPSS) collaboration searched for evidence of photons coupling to light, neutral bosons (and tested the particle interpretation of the PVLAS claim) in a series of measurements that took place at Jefferson Lab (JLab) in the Spring of 2007. The results reported here are the LIPSS collaboration's direct searches for the scalar coupling of photons to a hypothetical light neutral boson (LNB) in a regeneration experiment. This is contrasted to the search for pseudoscalar couplings between photons and a LNB that has already recently been reported by the BMV collaboration [16], and as carried out originally by the BFRT collaboration [17]. The GammeV collaboration at Fermi National Accelerator Laboratory (FNAL) [18] and the OSQAR collaboration at the European Center for Nuclear Research (CERN) [19] reported first results for both scalar and pseudoscalar couplings to photons in this same region of coupling-mass parameter space. It is to be emphasized that the LIPSS collaboration took data continuously over extended periods as compared with the previously reported LSW experiments. Thus, it is most sensitive to phenomena discussed in the context of hypothetical chameleon particles [20]. The limits presented here can also be compared with the results from the CAST collaboration [21] that searches for solar produced axions (light, weakly interacting, pseudoscalar bosons, [6]) using the regeneration technique, to sensitive searches for dark matter halo axions in the galaxy [22], and to constraints on BSM couplings and masses from tests of the gravitational inverse-square law [23].

The experimental setup is shown in Figure 2. Laser light from the JLAB Free Electron Laser (FEL) facility was used over a period of one week of running. The FEL creates light that is more than 99.9 % linearly polarized over a wide range of wavelengths in pulses that are 150 femtoseconds long with a 75 MHz repetition rate. For the LIPSS runs, it was tuned to a wavelength of $0.935 \pm 0.010$ micrometers with an intensity of 180 watts on average, and collimated to an 8 millimeter beam diameter; all parameters were monitored continuously during the run. The polarization direction of the laser light was verified with an optical polarization filter and chromo-color television cameras.

The beam exits the FEL optical transport system and is directed onto the LIPSS beam line through a series of water-cooled turning mirrors (TM's) and collimators, as shown in Figure 2. The TM's are specially coated to reflect 0.935 micron light and absorb light outside its narrow optical bandwidth (roughly 0.010 micrometers). The LIPSS beam line consists of an upstream (generation) magnetic field region and an identical (regeneration) magnetic field region placed downstream of it. Between the generation and regeneration magnets is an optical beam dump that also serves as a power meter; the beam dump in combination with a stainless steel vacuum flange on the input to the downstream beam line blocks all incident FEL light from the regeneration magnet. Any regenerated photons would be detected by the detector system housed in the Light Tight Box, downstream of the regeneration magnet.

Both generation and regeneration magnets had dipole fields of $1.77 \pm 0.04$ Tesla on average. Each magnet had an effect length of $1.01 \pm 0.02$ meters. The magnetic field direction was determined from the magnet pole configuration and verified using standard Hall probes. In all of the results presented here, the laser light polarization direction was perpendicular to the magnetic field; the experiment was therefore sensitive to scalar (positive parity) couplings between photons and LNBs.

The Light Tight Box in Figure 2 is an aluminum case painted on both inner and outer surfaces with black paint, housed inside a second box of black tape-covered aluminum foil. Inside the Light Tight Box, the photon beam passes a Newport KPX082AR16 50.2 millimeter lens which serves to focus the beam to desired accuracy onto the CCD array; the array sits five centimeters downstream of the lens. The camera system is a Princeton Instrument Spec-10: 400BR with WinView32 software. It consists of a back-illuminated CCD with 1340×400 pixels imaging area (a single pixel is 20 μm×20 μm in area) and a controller box for easy integrated measurement using a PC. The CCD array is cooled to –120º C resulting in a typical dark current of less than one single electron per pixel per hour [24]. The system featured onboard grouping (binning) of pixels, where groups of adjacent pixels may be summed before readout to decrease read noise. The detection system also consisted of a light emitting diode (LED) and a convex lens used to provide a beam spot on the CCD; this serves as a reference spot on the CCD.

High signal to noise ratios are needed to set the sensitive limits desired in this experiment. The noise in each pixel can come from a variety of sources: thermal, electronic, and stray light, to name a few. Additionally, a typical run may contain events due to cosmic rays (CRs) that strike the CCD array. All of these sources and others were studied and characterized over the past two years. Data were collected with the FEL on, with and without lasing, with both and either generation and regeneration magnets on and off, with the CCD camera shutter open and closed in each case. Stray light from fluorescence in gas in the vacuum pipe due to CRs was shown to be negligible since the experiment was run with $10^{-6}$ Torr. Stray light from all sources was shown to be less than one count per pixel per hour during the experiment. Read noise from the detector electronics was characterized in a series of short 'bias' runs of 0.01 seconds before, during, and after the data runs. The read noise was determined to be 2.7 ±0.2 counts per pixel per readout. The read noise was well described and easily subtracted from the data. Additionally, this contribution was minimized by collecting data for at least two hours in each run. CR's that strike the pixel array leave clear ionization signals in the pixels that they strike and are easily subtracted from the data. Runs that contain a CR hit on any pixel within an area of 100×100 pixels around the signal region were discarded [25].

The data were analyzed by defining a signal region where any regenerated photons would be observed, and background regions where no signal was expected. Light from a green (0.532 micrometers) laser placed upstream of TM1 was focused onto the CCD array through the focusing lens shown in Figure 2. Then, the FEL (in the so-called alignment mode where the laser average power was reduced by more than an order of magnitude so as not to damage the CCD optics) was aligned in precisely the same way and focused onto the array. In both cases, it was demonstrated that the laser light was focused by the lens down to the same, single pixel. Alignment mode runs were taken before and after the data runs, and were interspersed during the data runs in order to check for long term beam motion. No such effect was observed over the running period. The positions of the beam at TM1 to TM3 were monitored continuously during the data runs by cameras and Spiricon LBA-PC software. It was determined that the beam wandered by at most one centimeter over the two meter long beam line. This corresponds to less than 30 micrometers of displacement at the CCD array (which is 5 centimeters from the focusing lens). Thus, the signal region for the pixel array was taken to be a 3×3 pixel area at the lens focus. Tests performed subsequent to the data runs confirmed that the beam focus on the signal region wandered by at most one pixel vertically and horizontally; the 3×3 pixel area defined as the signal region did not change during the data runs.

The nine pixels in the signal area were binned together in software for each run. All other pixels and pixel groupings outside the signal region were used to define the background region(s). The difference between the counts in the signal region and the counts in the background region (normalized to the number of pixels in the signal region) was determined for all data runs [25]. Figure 3(a) shows the number of counts in the signal region (top plot) where any regenerated photons would register a count in the pixel array. No excess events above background (bottom plot in Figure 3(b)) were seen in any single run, or if all runs were combined. The background events are normalized to the same CCD array area in cases where a large area is use to get high background statistics. Seventeen hours of data were taken and analyzed.

The rate of regenerated photons is given by

$$R = r_\gamma P_{\gamma \to LNB} \, P_{LNB \to \gamma} \varepsilon_c \varepsilon_d \tag{1}$$

where $r_\gamma$ is the FEL (incident) photon rate, $\varepsilon_c$ is the photon collection efficiency (solid angle for detection), $\varepsilon_d$ is the detector quantum efficiency, and

$$P_{\gamma \to LNB} = P_{LNB \to \gamma} = \frac{(gB)^2}{\frac{m^4}{4\omega^2}} \sin^2\left(\frac{m^2 L}{4\omega}\right) \approx \frac{1}{4}(gBL)^2 \tag{2}$$

is the probability of scalar boson generation from the incident photons for magnets not too long for a given wavelength of light; photon regeneration from these scalar particles is given by the identical expression as shown. Here $\omega$ is the photon energy, $m$ ($g$) is the LNB mass (coupling strength to photons), and $B$ ($L$) is the magnetic field strength (length). The significance of the result is defined as $S = \text{signal}/\sqrt{\text{background}}$ where signal is the number of events expected based upon Equation (1) and that would show up in the signal region as described above and shown in Figure 3(a). Taking S greater than or equal to five as the criterion for a new discovery, the results indicate no coupling of photons to a LNB at this level.

The results from this run can therefore be used to set the new limits on the scalar coupling of photons to a hypothetical LNB shown in Figure 4. This represents the most stringent limits to date on this scalar coupling in a generation-regeneration experiment in this range of parameters for a long, continuously-running LSW experiment. The region above the *S*=5 curve (short dashed) and *S*=2 (full) is ruled out in the present experiment. These are similar limits already set by the BMV [16], BFRT [17], GammeV [18], and OSQAR [19] collaborations for pseudoscalar and scalar couplings, but under slightly different LSW experimental conditions. The limits set by the BFRT collaboration [17] (long dashed curve) is also presented in Figure 4.

The authors thank the technical staff of the Jefferson Lab Free Electron Laser Facility, especially F. Dylla, G. Neil, G. Williams, R. Walker, D. Douglas, S. Benson, K. Jordan, C. Hernandez-Garcia, and J. Gubeli, as well as M.C. Long of Hampton University for their excellent support of the experimental program. Funding from the Office of Naval Research Award N00014-06-1-1168 is gratefully acknowledged.

# References

1. M. Ahlers *et. al.*, Phys. Rev. D**75**, 035011 (2007); A. Ringwald, J. Phys. Conf. Ser. **39**, 197 (2006).
2. J. Khoury and A. Weltman, Phys. Rev. Lett. **93**, 171104 (2004), Phys. Rev. D **69**, 044026 (2004); D.F. Mota and D.J. Shaw, Phys. Rev D **75**, 063501 (2007), Phys. Rev. Lett. **97**,151102 (2006).
3. Ph. Brax, C. van de Bruck, A.-C. Davis, J. Khoury, and A. Weltman, Phys. Rev. D **70**, 123518 (2004); Ph. Brax, C. van de Bruck, A.-C. Davis, J. Khoury, and A.M. Green, Phys. Lett. B **633**, 441 (2006).
4. K. Choi, Phys. Rev. D **62**, 043509 (2000).
5. I. Waga and J. Frieman, Phys. Rev. D **62**, 043521 (2000); J. Frieman *et. al.*, Phys. Rev. Lett. 75, 2077 (1995); J.A. Frieman, C.T. Hill, and R. Watkins, Phys. Rev. D **46**, 1226 (1992).
6. R.D. Pecci and H.R. Quinn Phys. Rev. Lett. **38**, 1440 (1977); Phys. Rev. D **16**, 1791 (1977); S. Weinberg, Phys. Rev. Lett. **40**, 223 (1978); F. Wilczek, Phys. Rev. Lett. 40, 279 (1978).
7. G.G. Raffelt, in W.-M. Yao *et. al.*, (Particle Data Group), J. Phys. G **33**, 1 (2006).
8. P.W. Higgs, Phys. Lett. **12**, 132 (1964); G. Kane, J.F. Gunion, H.E. Haber, and S. Dawson, The Higgs Hunter's Guide (Addison-Wesley Pub. Co., Menlo Park), (1990) and references therein.
9. See, for example, LHC ATLAS collaboration, ATLAS Technical Design Report, (2006).
10. B. Holdom, Phys. Lett. B **166**, 196 (1986); R. Foot and X. G. He, Phys. Lett. B **267**, 509 (1991); K.R. Dienes *et. al.*, Nucl. Phys. B **492**, 104 (1997); S.A. Abel and B.W. Schofield, Nucl. Phys. B **685**, 150 (2004); L.B. Okun, Sov. Phys. JETP **56**, 502 (1982).
11. J. Jaeckel and A. Ringwald, Phys. Lett. B **659**, 509 (2008); M. Ahlers *et. al.*, Phys. Rev D **77**, 095001 (2008).
12. V.V. Popov and O.V. Vasil'ev Europhys. Lett. **15**, 7 (1991); D. Maity, S. Roy, and S. SenGupta, Phys. Rev. D **77**, 015010 (2008).
13. E. Zavattini *et. al*., Phys. Rev. Lett. **96**, 110406 (2006).
14. E. Zavattini *et. al.*, http://axion-wimp.desy.de/index_eng.html (2007).
15. K.V. Bibber *et. al.*, Phys. Rev. Lett. **59**, 759 (1987).
16. C. Robilliard *et. al.*, Phys. Rev. Lett. **99**, 190403 (2007).
17. R. Cameron *et. al.*, Phys. Rev. D **47**, 3707 (1993).
18. A.S. Chou *et. al.*, Phys. Rev. Lett. **100**, 080402 (2008); GammeV collaboration http://gammev.fnal.gov.
19. R. Ballou *et. al.*, CERN Report CERN-SPSC-2007-039 (2007); SPSC-M-762 (2007).
20. M. Ahlers *et. al.*, Phys. Rev. D **77**, 015018 (2008); H. Gies *et. al*., Phys. Rev. D **77**, 025016 (2008).
21. S. Andriamonje *et. al.* (CAST collaboration), J. Cosmol. Astropart. Phys. **04**, 010, (2007).
22. R. Bradley *et. al.*, Rev. Mod. Phys. **75**, 777 (2003); P. Sikivie, Phys. Rev. Lett. **51**, 1415 (1983).
23. G.L. Smith *et. al.*, Phys. Rev. D **61**, 022001 (2000).

24. LIPSS collaboration, to be published.
25. K.B. Beard, Jefferson Laboratory Technical Note, JLAB-TN-07-012 (2007).

**Figures**

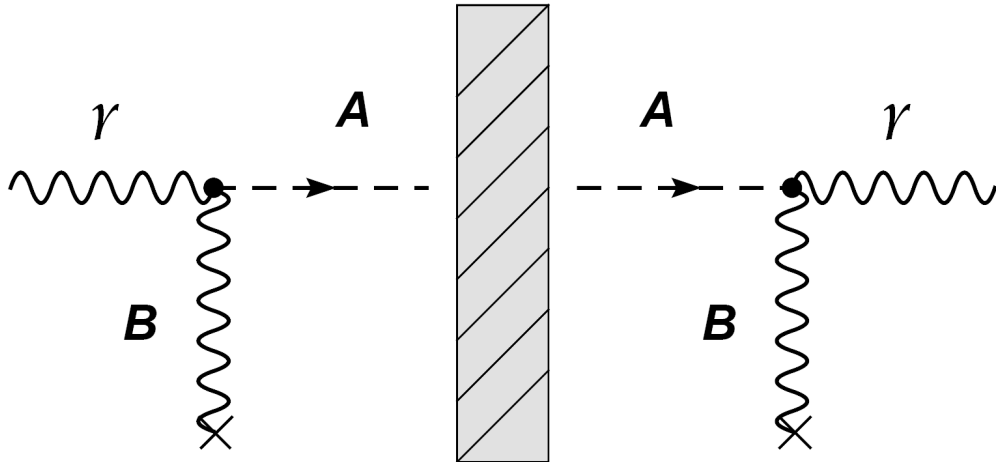

Figure 1. Photon regeneration using the "light shining through a wall" technique. The incident light (γ) couples to photons in the magnetic field (**B**) creating the hypothetical light neutral boson (**A**). Because it is weakly interacting, the LNB passes through the optical barrier (the "wall") while no incident photons do so. Regenerated photons having the same characteristics as the original photons result from the second magnetic field region downstream of the wall as shown.

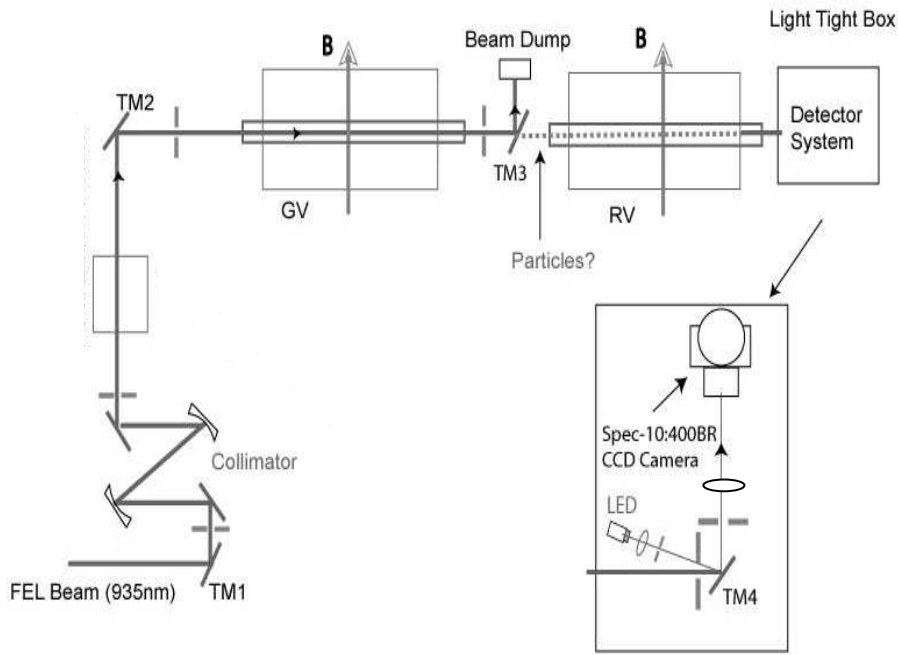

Figure 2. The LIPSS experimental setup. Laser light from the FEL is directed to the LIPSS beam line by Turning Mirror 1 (TM1) and a collimator. Turning Mirror 2 directs the incident light through the generation magnet (GV) and to an optical beam dump (the "wall") as shown. A second, identical magnet (RV) is used to regenerate any photons that would result from a hypothetical particle (a LNB) that passes through the wall; no incident FEL light passes into RV. These regenerated photons would be detected by the detector system in the Light Tight Box. Details of the Light Tight Box are shown in the insert.

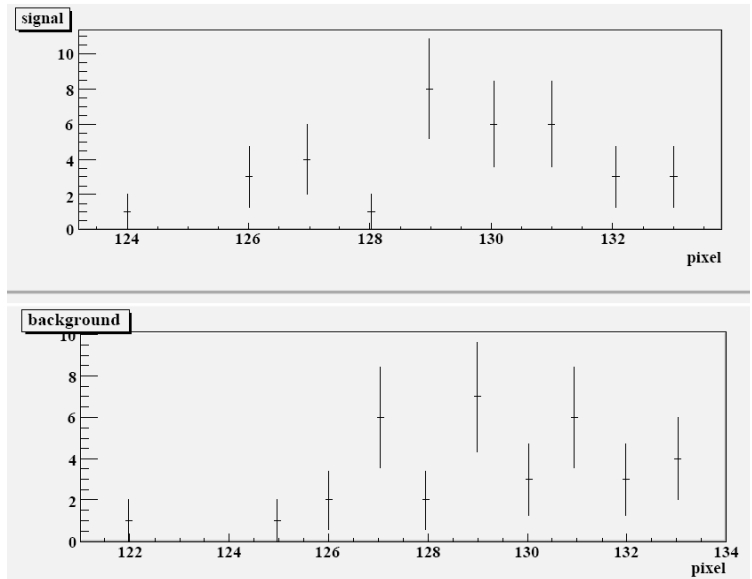

Figure 3. Data from the LIPSS runs in the signal region (3(a), top) and a representative background region (3(b), bottom) after proper normalization. The latter was used to determine the contribution from backgrounds in signal region. All pixels not in the signal region described in the text are used to determine the backgrounds. No excess of events above background is seen in the signal region.

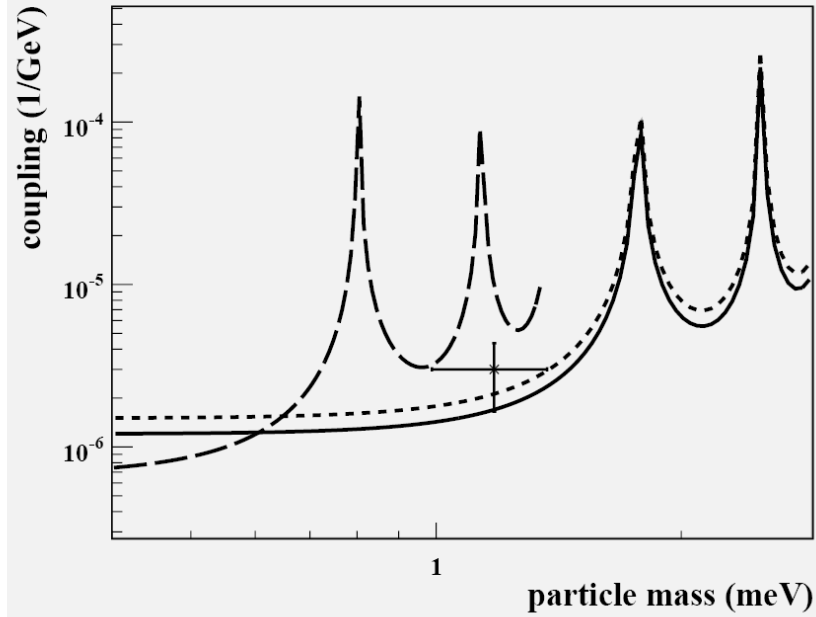

Figure 4. The new limits on scalar coupling of photons to a hypothetical LNB (in inverse giga-electron-volts) versus the LNB mass in milli-electron-volts. The curve shows the results for a significance of five (short dashed) and two (full). The BFRT result [17] is also shown (long dashed). The data point is the region claimed (now disclaimed) by the PVLAS collaboration [13].